\documentclass[12pt,preprint]{aastex}
\newcommand{\etal}{{et al.\ }}
\newcommand{\eg}{{e.g.,\ }}

\def\lsim{\mathrel{\rlap{\lower 4pt \hbox{\hskip 1pt $\sim$}}\raise 1pt\hbox {$<$}}}
\def\gsim{\mathrel{\rlap{\lower 4pt \hbox{\hskip 1pt $\sim$}}\raise 1pt\hbox {$>$}}}

\shorttitle{Off-axis properties of short gamma-ray bursts}
\shortauthors{Janka et al.}

\begin{document}

\title{Off-axis properties of short gamma-ray bursts}

\author{H.-Th.\ Janka\altaffilmark{1},
        M.-A.\ Aloy\altaffilmark{1,2},
        P.A.\ Mazzali\altaffilmark{1,3},
        and E.\ Pian\altaffilmark{1,3}} 
\email{thj@mpa-garching.mpg.de}
\email{maa@mpa-garching.mpg.de}
\email{mazzali@mpa-garching.mpg.de}
\email{pian@ts.astro.it}
\altaffiltext{1}{Max-Planck-Institut f\"ur Astrophysik,
       Karl-Schwarzschild-Str.\ 1, D-85741 Garching, Germany}
\altaffiltext{2}{Departamento de Astronom\'{\i}a y Astrof\'{\i}sica, 
       Universidad de Valencia, 46100 Burjassot, Spain}
\altaffiltext{3}{Istituto Naz. di Astrofisica - OATs, Via Tiepolo, 11,
       I-34131 Trieste, Italy}

\begin{abstract}
  
  Based on recent models of relativistic jet formation by thermal
  energy deposition around black hole-torus systems, the relation between
  the on- and off-axis appearance of short, hard gamma-ray bursts
  (GRBs) is discussed in terms of energetics, duration, average
  Lorentz factor, and probability of observation, assuming that the
  central engines are remnants of binary
  neutron star (NS+NS) or neutron star-black hole (NS+BH) mergers.
  As a consequence of the interaction with
  the torus matter at the jet basis and the subsequent expansion of
  the jets into an extremely low-density environment, the collimated
  ultrarelativistic outflows possess flat core profiles with only
  little variation of radially-averaged properties, and are bounded 
  by very steep lateral edges. Owing to the rapid decrease of the
  isotropic-equivalent energy near the jet edges, the probability of
  observing the lateral, lower Lorentz factor wings is significantly
  reduced and most short GRBs should be seen with on-axis-like
  properties. Taking into account cosmological and viewing angle
  effects, theoretical predictions are made for the short-GRB
  distributions with redshift $z$, fluence, and isotropic-equivalent energy.
  The observational data for short bursts with determined redshifts
  are found to be compatible with the predictions only if either 
  the intrinsic GRB rate density drops rapidly at $z \ga 1$, or a large
  number of events at $z > 1$ are missed, implying that the 
  subenergetic GRB~050509b was an extremely rare low-fluence event with
  detectable photon flux only because of its proximity and shortness.
  It appears unlikely that GRB~050509b can be explained as an off-axis
  event. The detection of short GRBs with small Lorentz factors is
  statistically disfavored, suggesting a possible reason for the absence
  of soft short bursts in the duration-hardness diagram. 

\end{abstract}


\maketitle

\section{Introduction}

The origin of short GRBs has long been the subject of theoretical
speculation.  A long-standing prediction is that short GRBs might
originate from NS+NS/BH mergers (\eg
\citealp{blinnikov84,Pa86,Ei89,Me02}, and references therein).
In the widely favored scenario a NS+NS/BH system gives rise to a BH
girded by a dense torus of gas from the disrupted NS companion. The BH
accretes mass at rates up to many solar masses per second, releasing
huge amounts of gravitational binding energy mostly in neutrinos,
gravitational waves, and through magnetohydrodynamic processes.

Hydrodynamic simulations (\citeauthor{AJM05}~\citeyear{AJM05}; AJM05)
show that thermal energy deposition around post-merger BH-torus
systems, \eg by neutrino-antineutrino annihilation, can drive
collimated, ultrarelativistic outflows with the high Lorentz factors,
internal variability and internal shocks that are deemed necessary to
explain GRBs with the fireball model (e.g., \citealp{Piran05}). The collimation
of observed short GRBs is rather uncertain and a matter of ongoing
studies. Evidence for collimation was claimed in case of GRB~050709
(\citealp{Foxetal05}), for which a jet semi-opening angle of
$\theta_\mathrm{jet}\sim 14^\mathrm{o}$ was inferred. 
Although more observations are needed to confirm the collimation of short
GRBs, this first indication seems to agree with the hydrodynamic simulations
of AJM05. If the majority of short bursts has opening angles of that size,
it would mean that only about 1\% of all bursts point to Earth.

Only in the past year have a few short GRBs been localised. Their properties
are recapped in Table~1. Two of these short GRBs are hosted by
elliptical galaxies, and also the host galaxy of GRB~051221 appears
to have a relatively evolved population of stars, despite its higher
star-formation rate. This adds confidence to the idea that the
progenitors of short GRBs are old systems. In addition, GRB~050509b is
in the outskirts of its potential host, in agreement with expectations
that coalescing compact binaries can travel large distances away from
their birth sites during their long gravitational-wave driven
inspiral \citep{Tutu94,BloSi99}. 
Moreover, a possible SN association has been ruled out for
GRB~050509b \citep{Hjorthetal05,ajct05} and for GRB~050709
\citep{Covinoetal05}.

Three out of five short GRBs were detected at similar redshifts ($z \sim 0.2$), 
but their intrinsic isotropic-equivalent energy release in gamma rays, 
$E_{\gamma,\mathrm{iso}}$, spans a factor of $\sim\,$100. The 
observation of intrinsically less energetic short GRBs at higher redshifts 
may be selected against. 

In this paper we use the jet models of AJM05 to discuss theoretical
predictions for the observable properties of short GRBs as a function
of viewing angle (Sect.~\ref{sec:jets}).  In
Sect.~\ref{sec:observability} we investigate the consequences of these
models for the probability of observing short bursts with different
isotropic equivalent energies and Lorentz factors, including the
selection effects due to cosmological redshift. A summary and
conclusions follow in Sect.~\ref{sec:conclusions}.

\section{Jets from post-merger black holes}
\label{sec:jets}

By means of relativistic hydrodynamics simulations AJM05 studied 
the formation of ultrarelativistic outflows from BH-torus systems. The BH
and torus masses and the deposition of thermal energy around the BH
were chosen as expected from NS+NS/BH merger models. For different
assumptions about the environment density and for varied geometry and
time-dependence of the energy deposition rate, AJM05 followed
the acceleration, collimation, and propagation of the
ultrarelativistic outflows for an evolution time of 0.5 seconds.  
  If the deposition rate of thermal energy per solid angle around the
  BH was sufficiently large, ultrarelativistic jets were launched
  along the rotation axis\footnote{A lower limit to the energy
    deposition rate needed to drive a relativistic outflow is set by
    the fact that the ram pressure of polar mass infall to the BH must
    be overcome (see AJM05 for details; model~B03 in that paper did
    not reach the energy deposition threshold and is therefore not
    considered here).}.  Although these
simulations are still far from taking into account all potentially
relevant physics and did not self-consistently track the
viscosity-driven torus evolution and its neutrino emission, they
nevertheless provide useful insight into the conditions and properties
of mass outflows from post-merger BH-torus systems.

The fact that an ultrarelativistic jet is launched from a ``naked'' BH-torus
system has important consequences for its properties. While in the case of
collapsars as sources of long GRBs the ultrarelativistic outflow originates
from BH-torus systems at the center of a massive star, the polar jets in 
NS+NS/BH mergers do not have to plough through many solar masses of
overlying stellar matter. Since the acceleration is not damped by swept-up
matter, the jets very quickly reach Lorentz factors of a few. The acceleration
is driven by the enormous radiation pressure of the pair-photon fireball
produced by the energy release around the BH. Collimation of the baryon-poor
jets is provided by the much denser torus walls which gird the evacuated polar
regions of the BH. As the jets propagate into the extremely low-density
environment, they continue to accelerate, reaching maximum Lorentz factors of a
few 100 within the simulated evolution time.

As a consequence of the interaction with the torus matter at the jet basis and
the subsequent free expansion, the collimated ultrarelativistic outflows
possess flat core profiles with only little variation of 
radially-averaged specific properties. 
These cores are bounded at their lateral edges by very steep gradients, which
are not smoothed by the prolonged entrainment of baryons as in case of
long GRB jets.  The rapid decrease of the isotropic-equivalent
energy as a function of $\theta$ implies, among other effects, 
that the probability of observing the lateral, lower Lorentz
factor wings must be expected to be significantly reduced.

Figure~\ref{fig:jetproperties} provides an overview of the
hydrodynamics results for the sample of ``type-B'' models discussed by
AJM05. In these simulations the BH-torus system was assumed to be
surrounded by a very rarified medium.
The models differ in the adopted rate of thermal energy deposition per
solid angle and in the prescribed time dependence. The parameters were
chosen such that the neutrino-antineutrino annihilation as calculated
from NS+NS/BH merger simulations, and in particular for the
post-merger accretion of a BH \citep{RJ99,Jetal99,JR02,SRJ04}, was
qualitatively and quantitatively reproduced.  Because of large
variations of the torus masses expected from compact object
mergers\footnote{The mass of the remnant torus varies with the masses
  and spins of the binary components and with their mass ratio; it is
  also sensitive to still uncertain properties of dense NS matter and
  depends on general relativity effects that are included only in a
  subset of the current NS+NS/BH merger simulations.}, AJM05 explored
a variety of energy deposition properties, approximately covering
the range of predictions from the above merger simulations (cf.\ Table~1 in
AJM05).

The corresponding differences in the assumed energy deposition
around the BH-torus systems account for the 
variation of the on-axis values of
the {\em isotropic-equivalent kinetic plus internal energies},
$E_\mathrm{iso}(\Gamma_\infty)$, of jet matter with estimated terminal
Lorentz factors $\Gamma_\infty > 100$ (Fig.~\ref{fig:jetproperties},
left). $\Gamma_\infty$ is estimated by assuming that 50\% of
the total specific energy will ultimately be converted to kinetic
energy at large radii\footnote{Note that an exact calculation of 
$\Gamma_\infty$ is beyond the scope of the hydrodynamic models of
AJM05, because the simulations do not include physics that plays 
a role during the later stages of the jet acceleration.}.
The models span a range of $\sim 100$ in $E_\mathrm{iso}$,
from some $10^{49}\,$erg to several 10$^{51}\,$erg. Depending on the 
efficiency of energy conversion to $\gamma$-ray emission, the
measurable $\gamma$-ray energy ($E_{\gamma,\mathrm{iso}}$) may 
be roughly a factor of 10 lower
than $E_\mathrm{iso}(\Gamma_\infty)$ (\eg \citealp{GSW01}, and
references therein).  All models show a flat core and steep lateral
wings in the radial average of $\Gamma$ 
(Fig.~\ref{fig:jetproperties}, right), but models
B07 and B08 have a somewhat more shallow decline of $E_\mathrm{iso}$
because a gradual decrease of the energy deposition rate was assumed
after an initial, burst-like phase. This caused a continuous decrease
of the jet opening angle, thus softening the wings of the jet profile.
It is currently not clear whether the jet-driving energy release of the 
BH-torus system follows such a burst/slow-decay behavior or whether it
is powerful and roughly constant over a longer period of time before it
ceases more abruptly, as assumed in the other models (B1--B6).

Figure~\ref{fig:jetproperties2} provides some observationally relevant
quantities deduced from the hydrodynamics results. The profiles of
$E_\mathrm{iso}(\Gamma_\infty)$ given in the left panel there
represent the isotropic-equivalent values of energy which is
potentially radiated from the outflow in different directions. These
profiles are calculated from the jet energies of
Fig.~\ref{fig:jetproperties}, taking into account radiation
contributions coming from regions outside of the line of sight.  
In order to compute these contributions, let us consider an
amount of energy ${\mathrm{d}}E'$ that is radiated into the solid angle
${\mathrm{d}}\Omega'$ at angle $\theta'$ relative to the
direction of motion (primed quantities are measured in the local
comoving frame). The transformation of the energy per solid angle
between the comoving frame and the laboratory frame (\eg a frame at
rest with respect to the central BH) is (\citealp{RL85})
\[
\frac{{\mathrm{d}}E}{{\mathrm{d}}\Omega}(\theta) = 
\frac{1}{\Gamma^3(1-\beta\cos{\theta})^3} 
                     \frac{{\mathrm{d}}E'}{{\mathrm{d}}\Omega'}\ ,
\]
where $\beta$ is the velocity of the comoving frame as measured in
the laboratory frame. Comparing the values of 
$\frac{{\mathrm{d}}E}{{\mathrm{d}}\Omega}$ at
two different angles, $\theta=\theta_1$ and $\theta=0$, one finds
that the energy emitted per unit of solid angle around an
angle $\theta_1$ can be expressed in terms of the energy radiated along 
the direction of propagation ($\theta=0$) as
\[
\frac{{\mathrm{d}}E}{{\mathrm{d}}\Omega}(\theta_1) =
\left(\frac{1-\beta}{1-\beta\cos{\theta_1}}\right)^3 
\frac{{\mathrm{d}}E}{{\mathrm{d}}\Omega}(0)\ ,
\]
if the radiation field is isotropic in the comoving frame. The emission
from the jet in a certain observer direction $\theta_1$ can now be obtained as 
the sum of the contributions from all matter moving in different directions 
with velocities $\beta_0$ at angles $\theta_0$ relative to the jet axis,
yielding
\begin{equation}
\left. \frac{{\mathrm{d}}E}{{\mathrm{d}}\Omega}(\theta_1)\right|_{\rm corr} =
\sum_{\theta_0} \left(\frac{1-\beta_0}{1-\beta_0\cos{(\theta_1 -
      \theta_0)}}\right)^3 \frac{{\mathrm{d}}E}{{\mathrm{d}}\Omega}(\theta_0)\ .
\label{eq:Ecorr}
\end{equation}
Since the profiles of $E_{\rm iso}(\theta)$ displayed in 
Fig.~\ref{fig:jetproperties} are related to the energy radiated per unit
solid angle into different directions $\theta = \theta_0$, the integration
given in Eq.~(\ref{eq:Ecorr}) can be directly applied to these profiles
in order to obtain the isotropic equivalent energy distributions
corrected for off-line-of-sight contributions. The results are plotted in
Fig.~\ref{fig:jetproperties2}. From the latter figure it is obvious
that energy release with $E_\mathrm{iso}(\Gamma_\infty) >
10^{48}\,$erg is confined to semi-opening angles between about
10$^\mathrm{o}$ and 20$^\mathrm{o}$, corresponding to a collimation
(``beaming'') factor of the two polar jets of $f_\Omega =
1-\cos\theta_\mathrm{jet}$ between 1.5\% and 6\%.

The relative observability of a short burst from a model of our
set at different polar angles $\theta$ is shown in
Fig.~\ref{fig:jetproperties2}, middle panel. We plotted there the
quantity
\begin{equation}
P(\theta)\, \equiv\, 
{\sin\theta\, E_\mathrm{iso}(\theta) \over
\int_{-1}^1 \mathrm{d}\cos\theta'\, E_\mathrm{iso}(\theta') } \ ,
\label{eq:probability}
\end{equation}
where as a rough estimate for the detectability of a burst we 
assumed $f_{\mathrm{det}}(\theta)\propto
E_\mathrm{iso}(\Gamma_\infty)$ (using for $E_\mathrm{iso}(\Gamma_\infty)$
the data from the left panel in Fig.~\ref{fig:jetproperties2}). The
probability distribution confirms the steep lateral edges visible in the other
quantities for models~B01--B06 and the slightly softer wings in models~B07 and
B08, which correlate with a sharp drop of the average Lorentz factor. The
probability $P(\theta)$ peaks at angles between 4$^\mathrm{o}$ and
$\sim\,$11$^\mathrm{o}$, which is a somewhat smaller range than that associated
with $E_\mathrm{iso}(\Gamma_\infty) > 10^{48}\,$erg. Morever, $P(\theta)$
nicely demonstrates the widening of the lateral visibility of a GRB with
increasing energy deposition per solid angle around the BH (cf.\ the
sequence of models~B02, B04, B06, B01, and B05).

The hydrodynamic jet models of
AJM05 track the formation of ultrarelativistic GRB-viable outflow. It
turns out that the ``shells'' ejected by the central engine, i.e. the
BH-torus system, accelerate much faster in the leading part of the outflow
than the shells in its lagging part. The rear shells need therefore a longer
time to reach velocities $v \approx c$. This different acceleration at
early and late times of the relativistic wind ejection leads to a
stretching of the overall radial length of the outflow, $\Delta$,
relative to the on-time $t_{\mathrm{ce}}$ of the central
engine\footnote{We define the on-time or activity time of the central
  engine as the period of time during which the energy release of the
  BH-torus system is sufficiently powerful to drive an
  ultrarelativistic outflow as required for GRBs.} times the
speed of light $c$, $\Delta > c t_{\mathrm{ce}}$ (see
Fig.~\ref{fig:stretching}). This stretching has the important consequence that
the overall observable duration of the GRB (in the source frame), $T =
t_\Delta = \Delta/c$, may be a factor of ten or more longer than the
activity time of the central energy source
(Fig.~\ref{fig:stretching}), even when the GRB is produced by internal
shocks. This is only seemingly in conflict with the
canonical fireball picture as discussed, \eg by
\cite{Piran05,KPS97,SP97,NP02}, who draw the conclusion that
``internal shocks continue as long as the source is active, thus the
overall GRB duration $T$ reflects the time that the inner engine is
active''.  It should in fact be noted that GRB light curve calculations in
these shell collision models are based on the assumption that the
shells move with {\em constant Lorentz factors and with a 
velocity very close to the speed of light}, which means that the
expansion of the shells is considered only {\em after the preceding
acceleration away from the central engine}.  Therefore the ``shell
emission time from the inner engine'' \citep{Piran05} or ``ejection
time'' \citep{KPS97} or ``activity time of the inner engine''
\citep{NP02} referred to in these works corresponds to the overall
length the fireball has attained {\em in the saturated
stage} after the initial acceleration. This time can only
be identified with the real activity time or on-time of the
central engine, whose energy output launches and drives the
initial shell ejection, if the shells accelerate to nearly the speed
of light within a negligibly small period of time. Only in this case
the claim of equal emission and source activity times applies and can
be traced back to the fact that a photon radiated from
a fireball shell is observed almost simultaneously with a
(hypothetical) photon launched at the central engine together with the
radiating shell \citep{NP02}. The hydrodynamic jet simulations, however,
show that the acceleration time, in particular for the shells ejected 
later, is not negligible (Fig.~\ref{fig:stretching}).

Taking the terminal length
$\Delta(\theta)$ of the ultrarelativistic outflow in direction
$\theta$ as a very crude measure of the overall observed 
duration $T(\theta)$ of
a burst\footnote{We estimate $\Delta(\theta)$ as the radial extension
  of those parts of the outflow which reach terminal Lorentz factors
  $\Gamma_\infty > 100$, where $\Gamma_\infty$ of a mass element in
  the outflow is again calculated by assuming that 50\% of its total
  energy will finally be converted to kinetic energy.}, we have
plotted this duration relative to the on-axis value in the right-hand
panel of Fig.~\ref{fig:jetproperties2}. The off-axis variations of the
possible GRB durations are sizable. Bursts which are seen more than
5$^{\mathrm{o}}$--10$^{\mathrm{o}}$ off-axis have only 30\%--60\% of
the on-axis duration.  The outflow stretching can easily be the dominant
effect determining the duration of short GRBs, but it is not possible to
assess its relevance for long bursts without hydrodynamical modeling.
This is due to the more complex and largely different dynamics of the
collimated relativistic outflow caused by the presence of the massive star
in the collapsar model. 

Predictions of the absolute
duration of the GRB emission based on jet simulations where 
the period of energy release was just a parameter are hard to make.
Self-consistent torus evolution models, including the energy production
mechanism and the feedback from jet formation, will be needed to provide
reliable numbers for the duration of the source activity and the 
differential acceleration of the ejecta shells at early and late times.
Moreover, the jet structure at 0.5$\,$s after the jet creation, at
which time AJM05 had to stop their calculations, may not be representative
for the situation at the time when the GRB is produced.

\section{Predicted short GRB distributions}
\label{sec:observability}

We can employ our set of jet models to attempt to predict
the observable short-GRB distributions versus redshift $z$, fluence,
isotropic equivalent energy output in $\gamma$-rays,
$E_{\gamma,\mathrm{iso}}$, and Lorentz factor $\Gamma$ of the flow
that produces the bulk of the measured $\gamma$-radiation. We discuss
these aspects in view of the observed bursts of
Table~1. Our methodology is different from that
of other groups, who used observational data of GRB redshift,
luminosity, and peak flux distributions to derive constraints on the
intrinsic properties (event rates or lifetimes, jet collimation,
luminosities) as functions of redshift (e.g.,
\citealp{Ando04,GP05,NGF05}). We instead refer to the sample of jet
models from AJM05 to define the intrinsic distribution of short
GRB energies as a function of viewing angle for the individual events.
This approach suffers from
the huge uncertainties of the theoretical models (see below)
and naturally can have only a tentative character.

In order to perform the analysis, we assume that our models have equal
probability and are representative of the intrinsic variability of the short
GRB source population at all redshifts. This is a crucial assumption and it 
can certainly
be questioned, but currently it can hardly be replaced by more realistic 
alternatives, because the uncertainties concerning the link between
NS+NS/BH binary parameters and GRB properties are still large.

In order to test the sensitivity of our analysis to a variation of the
comoving short GRB rate density as function of redshift,
$R_{\mathrm{GRB}}(z)$, we consider three cases, namely an intrinsic GRB 
rate which (i) is constant with comoving volume, (ii) follows the star
formation rate SF2$-$sfr of Eq.~(4) in \cite{GP05}, or (iii) varies
according to the binary neutron star merger rate, case SF2+delay of
Eq.~(6) in \cite{GP05}.

Based on these prescriptions, we produce the expected observable 
distributions of GRB properties by Monte Carlo sampling, randomly 
drawing GRB energies and Lorentz factors as functions of viewing angle 
$\theta$ from our set of GRB-jet models, using the results shown in 
Fig.~\ref{fig:jetproperties2} and Fig.~\ref{fig:jetproperties},
respectively. We estimate the isotropic-equivalent GRB energy output,
$E_{\gamma,\mathrm{iso}}(\theta)$, by
reducing the isotropic equivalent kinetic plus internal energy of the outflow,
$E_{\mathrm{iso}}(\theta)$ of Fig.~\ref{fig:jetproperties2}, 
by a factor of 10. This is supposed to 
account for the limited efficiency of energy conversion to 
$\gamma$-rays and the fact that only a fraction of the $\gamma$-emission
occurs in the energy band of a measurement\footnote{A careful
inclusion of a redshift-dependent k-correction appears inappropriate,
because we do not apply detailed theoretical arguments to estimate the 
conversion efficiency of jet energy to $\gamma$-radiation, and
model-dependent spectral properties of the $\gamma$-ray emission
are disregarded as well.}. Our set of models can be
considered to define a co-moving space distribution function 
$\widetilde{\Phi}(\theta,E_{\gamma,\mathrm{iso}},\Gamma)$, which is assumed
to be normalized to unity and whose integral over $\Gamma$ yields another
normalized distribution function, 
$\Phi(\theta,E_{\gamma,\mathrm{iso}})\equiv 
\int{\mathrm{d}}\Gamma\,\widetilde{\Phi}$. The random angle between the 
jet axis and the line of sight is denoted by $\theta$. 
Formally, the expected redshift distribution of measured bursts
can be computed from the intrinsic distribution as
\begin{equation}
\dot N(z_1,z_2) = \int\limits_{z_1}^{z_2}{\mathrm{d}}z\,
{{\mathrm{d}} V\over{\mathrm{d}}z}{R_{\mathrm{GRB}}(z)\over 1+z}
\int\limits_{-1}^{+1}{\mathrm{d}}\mu\,2\pi
\int\limits_{E_{\gamma,\mathrm{iso}}^{\mathrm{min}}(\theta,f_{\mathrm{min}},z)}
^{E_{\gamma,\mathrm{iso}}^{\mathrm{max}}(\theta)}
{\mathrm{d}}E\,\,\Phi(\theta,E)\ ,
\label{eq:eventsz}
\end{equation}
when $\dot N(z_1,z_2)$ represents the observed rate of events in the 
redshift interval $z_1 < z < z_2$ (in the following we will only
consider normalized distributions and the absolute value of the rate
will be of no concern). In Eq.~(\ref{eq:eventsz}), $\mu = \cos\theta$,
${\mathrm{d}} V/{\mathrm{d}}z = 4\pi D_L^2(z)c
[H_0(1+z)^2(\Omega_{\mathrm{M}}(1+z)^3 + \Omega_{\mathrm{K}}(1+z)^2 +
\Omega_{\Lambda})^{1/2}]^{-1}$ is the comoving volume element, and
$(1+z)^{-1}$ accounts for cosmological time dilation. 
The cosmological parameters used in our study are $H_0 =
72\,$km$\,$s$^{-1}$Mpc$^{-1}$, $\Omega_{\Lambda} = 0.72$,
$\Omega_{\mathrm{M}} = 0.28$, and
$\Omega_\mathrm{K}=1-\Omega_\mathrm{M} - \Omega_{\Lambda}$.  
In Eq.~(\ref{eq:eventsz})
as well as in Eqs.~(\ref{eq:eventsg})--(\ref{eq:eventse})
below, an instrument specific detection probability that depends
on the photon flux and thus on the burst luminosity, spectrum, and
redshift, is ignored (we have no information on this quantity in
terms of the fluence). 
$E_{\gamma,\mathrm{iso}}^{\mathrm{min}}(\theta,f_{\mathrm{min}},z)$ is
the isotropic equivalent energy corresponding to the minimum fluence 
$f_{\mathrm{min}}$ that can be measured by the detector, and 
$E_{\gamma,\mathrm{iso}}^{\mathrm{max}}(\theta)$ corresponds to the
maximum energy release of the models of our sample in a 
certain direction $\theta$. The fluence
at the instrument is computed from the intrinsic energy output
$E_{\gamma,\mathrm{iso}}(\theta)$ (in the reference frame of the 
source) and from the luminosity distance
$D_L(z)$ as $f(\theta,z) = (1+z)\,E_{\gamma,\mathrm{iso}}(\theta)/
[4\pi D_L^2(z)]$.

The predicted distribution of Lorentz factors of the observed
bursts can be written as
\begin{equation}
\dot N(\Gamma_1,\Gamma_2) = \int\limits_{-1}^{+1}{\mathrm{d}}\mu\,2\pi
\int\limits_{\Gamma_1}^{\Gamma_2}{\mathrm{d}}\Gamma
\int{\mathrm{d}}E_{\gamma,\mathrm{iso}}\,
\widetilde{\Phi}(\theta,E_{\gamma,\mathrm{iso}},\Gamma)
\int\limits_{0}^{z_{\mathrm{max}}
(E_{\gamma,\mathrm{iso}}(\theta),f_{\mathrm{min}})}{\mathrm{d}}z\,
{{\mathrm{d}} V\over{\mathrm{d}}z}{R_{\mathrm{GRB}}(z)\over 1+z}\ ,
\label{eq:eventsg}
\end{equation}
where $z_{\mathrm{max}}(E_{\gamma,\mathrm{iso}}(\theta),f_{\mathrm{min}})$
is the maximum redshift at which GRBs with isotropic equivalent energy
$E_{\gamma,\mathrm{iso}}(\theta)$ produce a fluence above the lower
detection bound $f_{\mathrm{min}}$. Similarly, the fluence distribution
is given by
\begin{equation}
\dot N(f_1,f_2) = \int\limits_{-1}^{+1}{\mathrm{d}}\mu\,2\pi
\int\limits_{0}^{\infty}{\mathrm{d}}z\,
{{\mathrm{d}} V\over{\mathrm{d}}z}{R_{\mathrm{GRB}}(z)\over 1+z}
\int\limits_{E_{\gamma,\mathrm{iso}}(f_1(z))}^{E_{\gamma,\mathrm{iso}}(f_2(z))}
{\mathrm{d}}E\,\Phi(\theta,E)
\label{eq:eventsf}
\end{equation}
with $f_2 > f_1 \ge f_{\mathrm{min}}$, when $E_{\gamma,\mathrm{iso}}(f_1(z))$
and $E_{\gamma,\mathrm{iso}}(f_2(z))$ are the isotropic equivalent 
energies of $\gamma$-radiation that account for fluences $f_1$ and 
$f_2$ in the frequency window of the detector for a GRB at redshift $z$. 
The distribution versus $E_{\gamma,\mathrm{iso}}$ can be represented by
\begin{equation}
\dot N(E_{\gamma,\mathrm{iso}}^1,E_{\gamma,\mathrm{iso}}^2) = 
\int\limits_{-1}^{+1}{\mathrm{d}}\mu\,2\pi
\int\limits_{E_{\gamma,\mathrm{iso}}^1}^{E_{\gamma,\mathrm{iso}}^2}
{\mathrm{d}}E_{\gamma,\mathrm{iso}}\,
\Phi(\theta,E_{\gamma,\mathrm{iso}})
\int\limits_{0}^{z_{\mathrm{max}}
(E_{\gamma,\mathrm{iso}}(\theta),f_{\mathrm{min}})}{\mathrm{d}}z\,
{{\mathrm{d}} V\over{\mathrm{d}}z}{R_{\mathrm{GRB}}(z)\over 1+z}\ .
\label{eq:eventse}
\end{equation}

We perform our investigation with two different values of the lower
fluence cutoff for detection (i.e., the detection threshold is assumed
to be a step function). For one simulation we use $f_{\mathrm{min}}=
10^{-8}\,$erg$\,$cm$^{-2}$, which corresponds to the fluence of the
very weak GRB~050509b and thus can be considered as a lower bound of
the sensitivity of Swift. In a second simulation we use a cutoff value of
$f_{\mathrm{min}}= 1.6\times 10^{-7}\,$erg$\,$cm$^{-2}$, which we
derive from the limiting flux, $\phi_{\mathrm{min}} = 
f_{\mathrm{min}}/[(1+z)T_{\mathrm{i}}\epsilon_\gamma]$, 
of 1$\,$photon$\,$cm$^{-2}$s$^{-1}$ 
adopted for the BATSE instrument by \cite{GP05}, making the assumption
that $(1+z)T_{\mathrm{i}}\epsilon_\gamma = 100\,$keV$\,$s is a 
representative value for the product of photon detection time
interval and energy ($T_{\mathrm{i}}$ is the intrinsic duration 
of the burst). This higher fluence threshold seems to be
compatible with the sample of (bright) BATSE bursts listed in Table~1 of
\cite{GGC04}. We note that Nakar \etal(\citeyear{NGF05}) employ a
detection threshold of 1$\,$photon$\,$cm$^{-2}$s$^{-1}$ also for
Swift. A higher value for $f_{\mathrm{min}}$ may further be 
motivated by the fact that the photon flux decreases with increasing
redshift more rapidly than the fluence by an additional factor 
$(1+z)^{-1}$. Therefore a low fluence threshold for detection may 
overestimate the number of high-$z$ events observed.

To obtain normalized distributions, $\dot N(x_1,x_2)/\dot N$, 
we do not evaluate the
integrals in Eqs.~(\ref{eq:eventsz})--(\ref{eq:eventse}) directly,
but perform Monte Carlo sampling of a large number of GRB events at different
redshifts, with random orientations and with properties drawn
randomly from the model data plotted in Figs.~\ref{fig:jetproperties} and
\ref{fig:jetproperties2}, 
thus constructing the discrete distribution functions
$\widetilde{\Phi}$ and $\Phi$ which describe the
GRB properties according to our set of jet simulations. The sampled events
with $f \ge f_{\mathrm{min}}$ are then collected into bins to build up
the normalized probability distributions versus redshift,
fluence, $E_{\gamma,\mathrm{iso}}$, and $\Gamma$ shown in 
Fig.~\ref{fig:grbobservability}. We point out that the Lorentz
factors are based on those of Fig.~\ref{fig:jetproperties} and
therefore reflect the situation only 0.5$\,$s after the ejection
of relativistically expanding matter. Because of subsequent acceleration
and conversion of internal to kinetic energy (which we are unable to
trace in the hydrodynamic simulations), the terminal values
$\Gamma_\infty$ must be expected to be larger by a factor of 2--3.

Our predictions for the observable distributions vs.\ redshift, fluence,
energy, and Lorentz factor for the two chosen values of the fluence
cutoff are displayed in Fig.~\ref{fig:grbobservability}.  The
corresponding locations of four of the five bursts listed in
Table~1 are also indicated. The very bright short
GRB~051221 is outside of the energy scale of the third panel, hence
we omitted it from 
Fig.~\ref{fig:grbobservability}.  This burst is too energetic to be
accounted for by the range of $\gamma$-burst energies considered in
our set of jet models, in which we assumed that 10\% of the ejecta
energy can be converted to radiation in the frequency band of the
measurement.  But if this fraction were significantly higher --- the
observations of GRB~051221 indeed suggest a total efficiency of
60--70\% (\citealp{Soderbergetal06}) --- then the output of 
$\gamma$-energy
of this burst would be within the reach of our models for near-axis
observation (cf.~Fig.~\ref{fig:jetproperties}).  Our jet models
invoked energy deposition rates as obtained for the annihilation of
neutrino-antineutrino pairs in NS+NS/BH merger and post-merger
accretion simulations (for details and references to original work,
see AJM05). We therefore conclude that GRB~051221, if it is indeed
located at a redshift of 0.546, does not make a strong case for
different underlying energy extraction mechanisms (\eg
magnetohydrodynamic processes) and/or different progenitors as
speculated by \cite{Soderbergetal06}.

It is obvious that the choice of the lower fluence cutoff has a big
impact on the redshift distribution (left panels). For a detection
threshold of $10^{-8}\,$erg$\,$cm$^{-2}$ and our employed 
comoving GRB rate densities, the majority of bursts should
be seen at redshifts $z > 1$, while with the higher cutoff we predict
that bursts with $z > 0.75$ would not be detected. Clearly, the
observations of the recent bursts with redshifts between 0.16 and 0.72
are more compatible with the latter case. At the request of one referee,
we performed a Kolmogorov-Smirnov (KS) test
(not including the very energetic GRB~051221), which gives 
significance levels of 5\%, 8\%, and 20\% for the
consistency of the observations with the theoretical distributions for
the three tested GRB rate densities SF2$-$sfr, SF2+delay, and
``constant'', respectively (the KS probabilities in case of the lower
cutoff value are 0.1\%, 0.4\% and 2\%). Better agreement is
therefore found when the simulation predicts a larger number of nearby
events. Highest preference is attributed to the constant GRB rate
density, in which the relative fraction of visible bursts at $z < 0.5$
is largest.  This is in qualitative agreement with the recent analysis
by Nakar \etal(\citeyear{NGF05}; see Fig.~5 there), who found that
the well-localized bursts imply a high local rate of short-hard GRBs
events, and that long lifetimes of the progenitor systems are favored.
Given the small number of
events, the results clearly will change significantly with every new
observation (see \citealp{BP06}).

The fluence distributions show the expected increase towards
faint events, and the distribution of $E_{\gamma,\mathrm{iso}}$  
reveals a strong bias towards events with higher isotropic-equivalent
energies, corresponding to near-axis observation. Such events are 
much more likely to yield a fluence above the threshold values at 
large redshifts and are also preferred because of the steepness of the
jet wings, which reduces the probability of seeing events from those wings.
GRB~050509b is an exceptionally weak event, which is well separated
from the other observed bursts, and which the theoretical distributions
do not account for. It thus distorts the KS measures for the
fluence and energy distributions, which otherwise signal reasonably
good compatibility between calculations and observations, in particular
for the higher fluence threshold.
GRB~050509b is most probably {\em not} an off-axis observation, but 
requires either an event with lower energy output than for any of the
jet models in our sample, or an efficiency of the energy conversion to
$\gamma$-rays lower than assumed here. The latter is disfavored
by the observations of \cite{Bloometal05}. These authors also
discard GRB~050509b as an off-axis event on the basis of the very
early decay of the afterglow light curve. 

The probability distribution of the Lorentz factor reflects the
discreteness of our sample of models, because the redshift has no
influence on $\Gamma$ and because the jet wings are extremely steep
and the off-axis visibility much reduced (see
Fig.~\ref{fig:jetproperties}). Therefore, the high-energy model~B05
with a relatively wide jet clearly sticks out in the
$\Gamma$-distribution. A continuous set of models would enhance and
broaden this peak, and the ``underwood'' of events extending to low
values of $\Gamma$ would even be more reduced if terminal Lorentz
factors $\Gamma_\infty$ instead of $\Gamma$ were plotted. 

We hypothesize that this clear bias towards observing high-$\Gamma$
events may be connected with the lack of short-soft GRBs in the
duration-hardness diagram. Short bursts are on average harder than
long bursts, and short-soft bursts are rare or missing, because short
GRBs develop jets with higher Lorentz factors and very steep edges.
The underlying reason behind this effect is the fact that jets from
post-merger BH-torus systems do not have to make their way out of a
massive star through layers of dense stellar material, which can damp
the outflow acceleration by mass entrainment. In the commonly used
internal shock scenario for the GRB emission the peak energy of the
(synchrotron) spectrum, $E_{\rm p}$, obeys the relation $E_{\rm p}
\propto \Gamma^{b}$ with $b\sim -2$ (\eg \citealp{DM03,RL02,ZM02}).
In view of our jet models this would imply that short bursts might
have lower $E_{\rm p}$ but nevertheless be harder than long ones
because of a steeper increase of the $\nu F_\nu$ spectrum below the
peak.  The results of Ghirlanda \etal(\citeyear{GGC04}) seem to confirm 
this. Their spectral analysis for a sample of short bright GRBs
detected by BATSE in comparison with the spectral properties of long
bright BATSE bursts indeed reveals a significantly harder (i.e., less
negative) spectral index $\alpha$ rather than a larger peak energy
$E_{\rm p}$ for short GRBs. This leads to an increased hardness ratio
if $E_{\rm p}$ is above the higher energy band of the spectral
hardness measure, or overlaps with it, which is the case for almost
all of the short GRBs listed in Table~1 of Ghirlanda
\etal(\citeyear{GGC04}). We point out that the difference between the
peak spectral energies of short and long bursts might even be more
pronounced than found by Ghirlanda \etal(\citeyear{GGC04})
if long bursts occurred at typically higher redshifts and the intrinsic
values of these energies were compared.

However, the discussion by Zhang \& M{\'e}sz{\'a}ros (2002) also 
leaves possibilities how in the internal shock model
higher Lorentz factors of short GRBs 
may be compatible with {\em higher} peak spectral energies.
In case of Poynting-flux dominated outflow, Zhang \& M{\'e}sz{\'a}ros 
(2002) expect a direct proportionality, $E_{\rm p} \propto \Gamma$ 
(see their Eq.~21 and case (b) in their Fig.~3),
and for a kinetic-energy dominated fireball they find
$E_{\rm p} \propto L^{1/2} R_{\mathrm{int}}^{-1}$ (see their Eq.~17)
when $R_{\mathrm{int}} \approx \Gamma^2 c \delta t$ is the 
radius where internal shock collisions produce $\gamma$-ray emission
(see, e.g., Piran 2005). Although short GRBs have higher $\Gamma$'s
compared with long GRBs, their variability timescale $\delta t$
might be generically much smaller. This would result in a smaller
shock dissipation radius and therefore in a higher $E_{\rm p}$, 
because magnetic fields may be stronger at a smaller radius.

\section{Discussion and Conclusions}
\label{sec:conclusions}

Using a set of hydrodynamic simulations of ultrarelativistic jet
formation by thermal energy release around BH-torus systems as
remnants of NS+NS/BH mergers (AJM05), we have investigated the
off-axis visibility of short GRBs produced by such outflows.  The jets
are characterized by narrow cores with high isotropic-equivalent
energies and very high Lorentz factors, which are laterally bounded by
steep wings. These properties are a consequence of the specific
conditions in which the jets are considered to be launched, i.e.,
their acceleration away from the torus-girded BH, where the energy is
released, into an environment of extremely low density.  The
observability of short GRBs is therefore strongly favored within a
cone of semi-opening angle between about 10$^{\mathrm{o}}$ and
15$^{\mathrm{o}}$ around the jet axis.

We argued that the observable duration of short GRBs both in the
external shock model and in the
internal shock collision model for GRB production can be
significantly longer than the activity time of the central engine
whose energy release powers the GRB-viable ultrarelativistic outflow.
The reason for this claim is the fact that our jet formation
simulations showed that during the acceleration phase the fireball
experiences significant radial stretching, because hydrodynamic
effects cause the acceleration to proceed differently in the leading
and lagging parts of the outflow.  Therefore the overall radial length
$\Delta$ of the relativistic ejecta can become a factor of ten or more
larger than $c\,t_{\mathrm{ce}}$, when $t_{\mathrm{ce}}$ is the
on-time of the central engine (Fig.~\ref{fig:stretching}). This seems
to be in conflict with the canonical internal shock scenario,
according to which the observed GRB is produced by shell collisions
such that the observed light curve reflects the temporal activity and
overall duration of the energy release by the ``inner engine''
\citep{Piran05}. However, the standard picture applies only if
the fireball shells move immediately after their ejection with
velocities that are very similar (but
not necessarily identical) and very close to the speed of light.
This can be traced back to the fact that a photon radiated from
a fireball shell is observed almost simultaneously with a
(hypothetical) photon emitted from the central engine together with the
radiating shell \citep{NP02}.  If the acceleration is differential,
and in particular if the acceleration of shells ejected by
the central engine at later times proceeds more slowly, the fireball
evolution after creation at the engine is more complex
(Fig.~\ref{fig:stretching}). Its emission pattern then does not
directly replicate the activity timescales and the on-time of the central
source.

\begin{figure}
\plotone{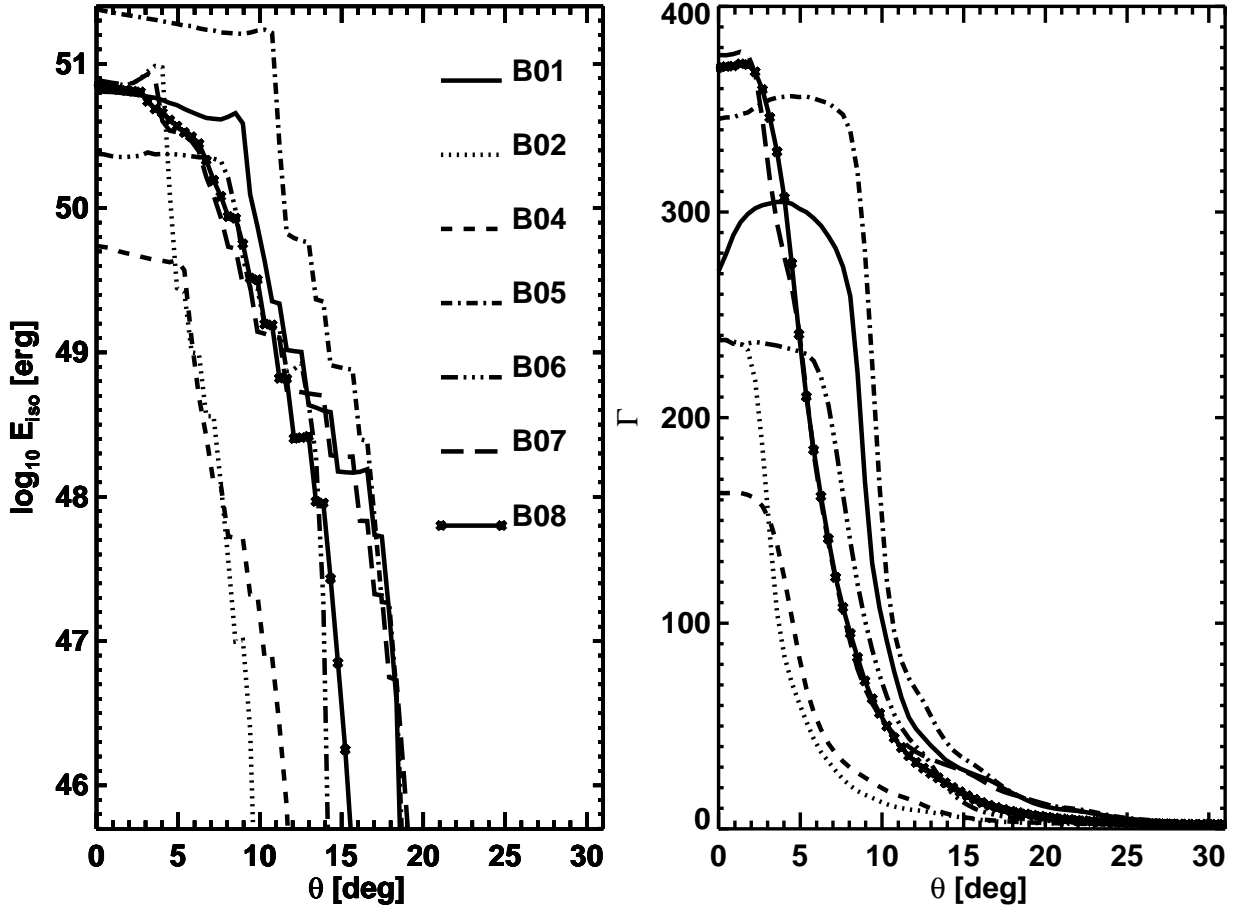}
 \caption{Properties of relativistic outflows from BH-torus systems
   as functions of viewing angle, determined by hydrodynamic
   simulations which show that collimated, ultrarelativistic mass ejection
   can be driven by sufficiently powerful deposition of thermal energy
   around the relic BH-torus systems of NS+NS/BH mergers.
   models~B01--B08 differ in the rate of energy deposition per unit 
   solid angle around the BH (AJM05).  
   {\em Left:} Total (internal
   plus kinetic) isotropic equivalent energy of matter
   with terminal Lorentz factors $\Gamma_\infty > 100$. 
   This energy may be larger than the GRB energy by a factor
   depending on the efficiency of conversion of outflow energy to
   $\gamma$-rays. 
   {\em Right:} Radially
   averaged Lorentz factor $\Gamma$ at 0.5 seconds after the onset of
   energy deposition (AJM05).  Note that the terminal value
   $\Gamma_\infty$ can be higher by a factor of 2--3.  
\label{fig:jetproperties}}
\end{figure}

\begin{figure}
\plotone{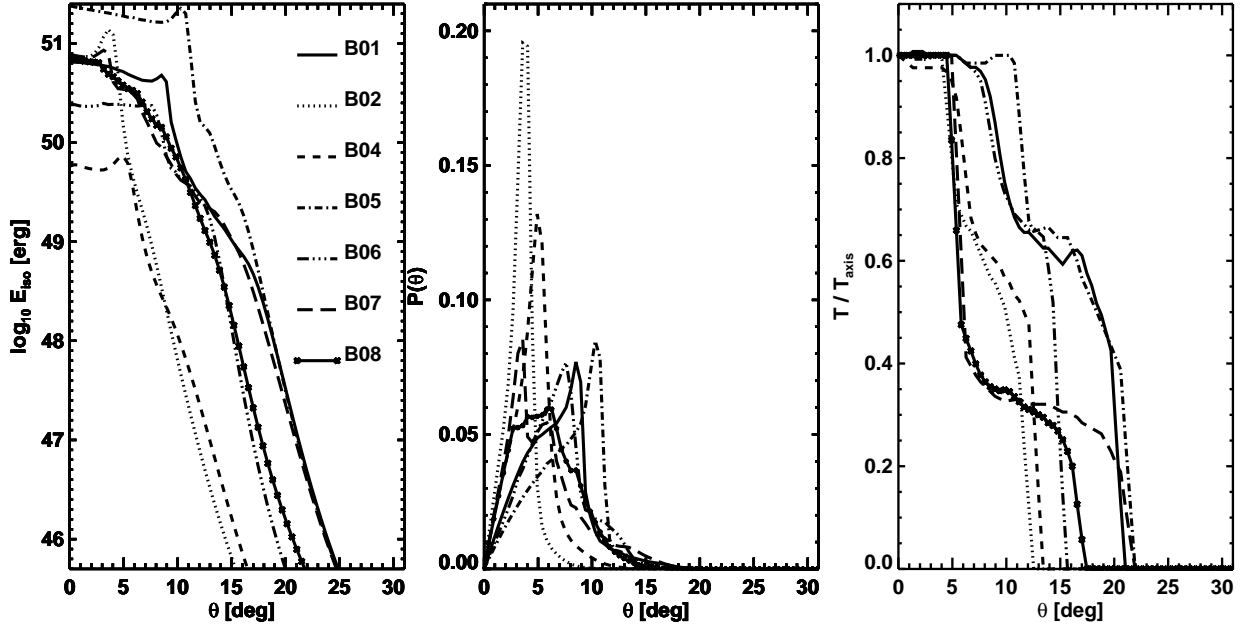}
 \caption{Observationally relevant properties of relativistic outflows 
   from BH-torus systems
   as functions of viewing angle, calculated using the results 
   of the hydrodynamic models of Fig.~\ref{fig:jetproperties}.
   {\em Left:} Total energy potentially radiated from ultrarelativistic,
   collimated ejecta with terminal Lorentz factors $\Gamma_\infty >
   100$, including the contributions from matter moving not along
   the line of sight (see text for details). The main difference
   compared to the left panel in Fig.~\ref{fig:jetproperties} is a 
   smoothing of the jet wings and a slight widening at energies 
   below about $10^{49}\,$erg.
   {\em Middle:} Relative observability of short GRBs from
   different viewing angles, estimated according to
   Eq.~(\ref{eq:probability}), using the $E_{\mathrm{iso}}$-values
   shown in the left panel.
   {\em Right:} Timescale of the observable GRB, estimated from the
   polar-angle dependent radial length $\Delta(\theta)$ of the 
   ultrarelativistic outflow with $\Gamma_\infty > 100$ in
   the observer frame, relative to the on-axis duration.
\label{fig:jetproperties2}}
\end{figure}

\begin{figure}
\plottwo{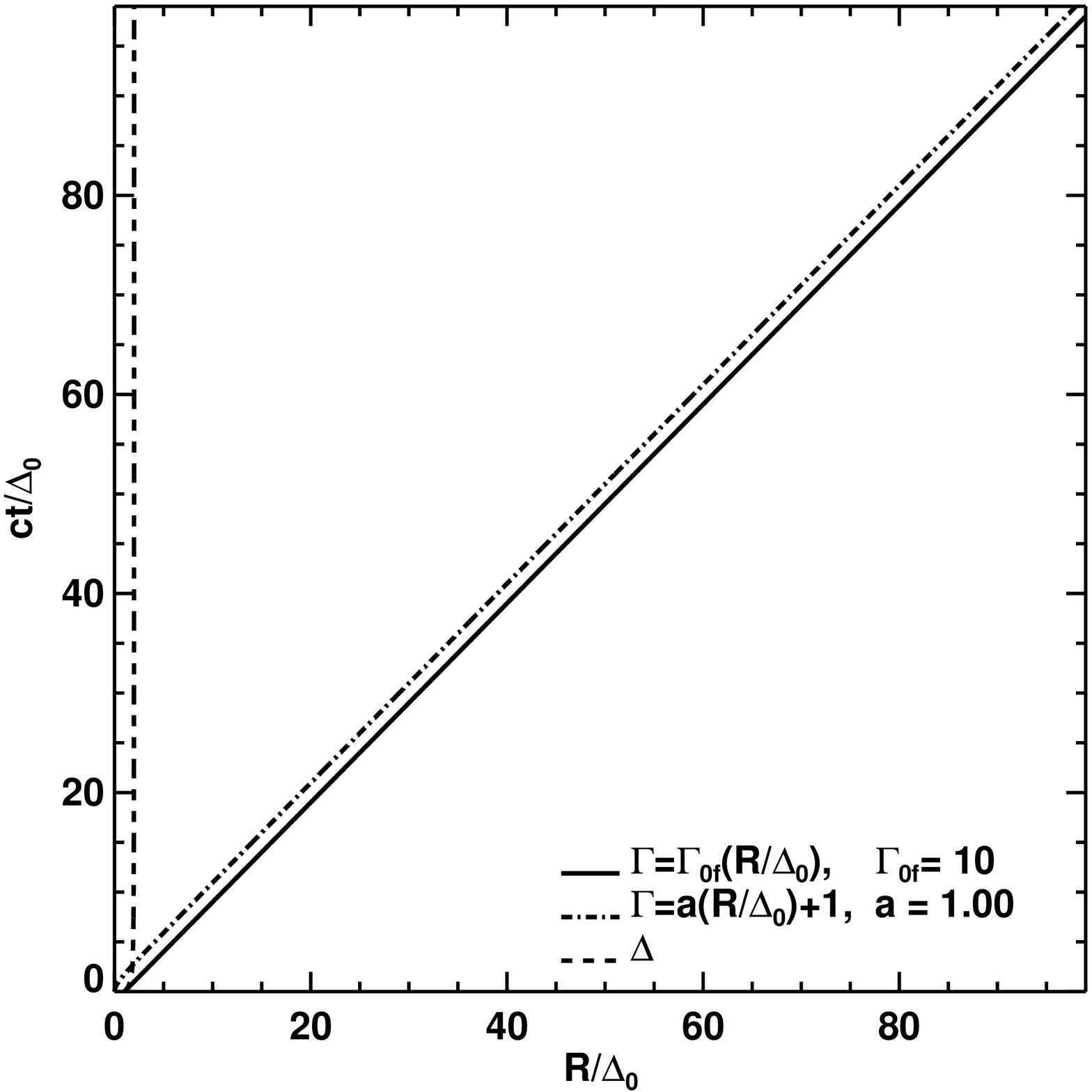}{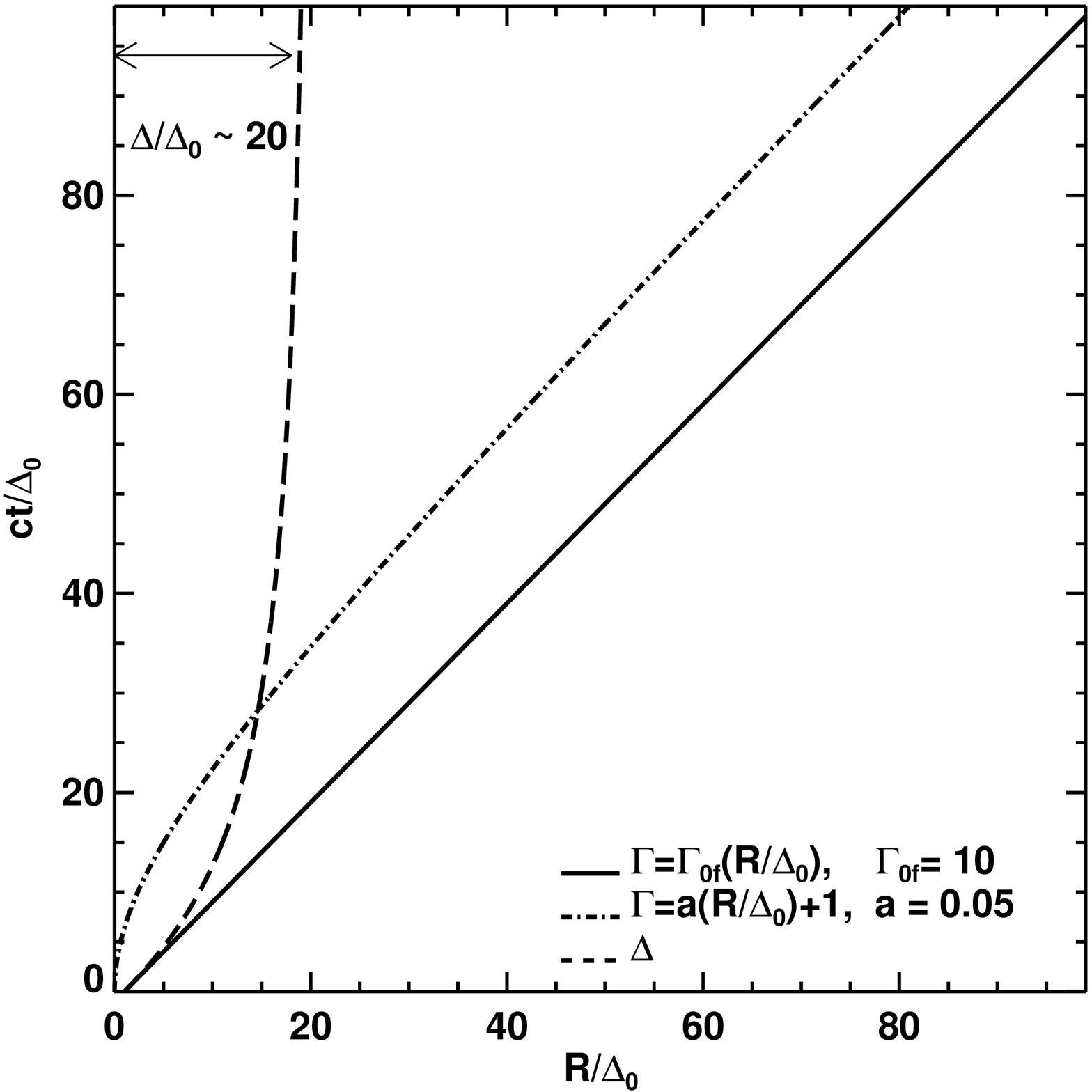}
 \caption{World-line diagrams for the evolution of shells at the 
   forward and rear edges of the ultimately highly relativistic
   ejecta. While in the plot on the left the Lorentz factor is assumed to
   increase linearly and very quickly with distance from the source for both
   the leading and lagging shells, the plot on the right displays a situation
   that fits better the results of the hydrodynamic simulations of jet
   formation: The lagging shell accelerates more slowly than the leading
   one. The acceleration laws are indicated in the lower right corners
   of the two panels. Time $t = 0$ is the moment when the shell
   at the rear edge of the ejecta is born, and $\Delta_0 =
   c\,t_{\mathrm{ce}}$ is the radial extension of the outflow at this
   time, when $t_{\mathrm{ce}}$ is the period of activity of the
   central engine. All lengths are normalized to $\Delta_0$. In
   the figure on the left the radial thickness of the expanding shell nearly
   saturates at a value of about 2$\Delta_0$, whereas on the right
   one can see a stretching of the fireball compared to its
   initial length by roughly a factor of 20, before a much slower
   growth indicates a nearly saturated state. The parameters 
   for this plot were chosen to closely reproduce the results of one
   of the jet models of AJM05.
\label{fig:stretching}}
\end{figure}

In order to investigate the consequences of the reduced off-axis visibility of
short GRBs produced by our jets, we made an attempt to predict
observable short-hard GRB distributions with redshift, fluence,
isotropic-equivalent energy, and Lorentz factor.  For this purpose we
made use of a sample of seven ultrarelativistic jet models
(AJM05), which roughly covers the range of energy release expected from
neutrino-antineutrino annihilation in NS+NS/BH binary mergers.
Employing these models for defining the intrinsic GRB properties, we
performed a Monte Carlo sampling of the expected observed event
distributions. Our analysis took into account the off-axis variation
of the jet properties as provided by the hydrodynamic simulations
(Figs.~\ref{fig:jetproperties} and \ref{fig:jetproperties2}) and referred
to standard prescriptions for the intrinsic event rate density as a 
function of redshift (constant rate as well as the cases SF2$-$sfr and
SF2+delay from \citealp{GP05}).  We found that the resulting distributions
predict far too many bursts at $z > 1$ and therefore show no good
agreement with the recent observations when a fluence threshold of 
$10^{-8}\,$erg$\,$cm$^{-2}$ as suggested by GRB~050509b was adopted
(the KS probability for the observed and predicted data being drawn from
the same distribution is less than 1\%). Improved consistency was
achieved when we performed our analysis with a higher fluence threshold
of $1.6\times 10^{-7}\,$erg$\,$cm$^{-2}$. In this case the bursts at
$z \ga 0.8$ escape from observation and the KS significance increases
to $\sim\,$20\% for the constant comoving GRB rate density. This would
imply that GRB~050509b was a very special and extremely rare case with
an exceptionally low fluence, which due to its shortness and proximity
still had a sufficiently large photon flux of about 
1$\,$photon s$^{-1}\,$cm$^{-2}$ and was therefore above the Swift 
detection threshold. Comparison of the results for different comoving
GRB rate densities revealed that the consistency between theory and
observations increases when the rate yields a relatively larger number
of events at redshifts $z \la 0.5$ rather than an increase for higher
values of $z$. Our results agree
qualitatively with those of Nakar \etal(\citeyear{NGF05}), who found 
that the detection of the current sample of low-$z$ short GRBs
is best compatible with a very low comoving rate density 
for $z > 1$ and a peak at $z \la 0.5$. 

The extremely weak GRB~050509b is clearly separated in energy
from the other
observed events and cannot be accounted for by off-axis observation of
any of the modelled jet outflows. It seems to require an intrinsically
less energetic event, because an efficiency of energy conversion to
$\gamma$-rays significantly lower than what we used is
disfavored by observations \citep{Bloometal05}. On the other hand, the
very energetic GRB~051221 exceeds the isotropic-equivalent GRB
energies predicted by our jet models, making the assumption that
10\% of the ejecta energy can be converted to radiation in the
detector frequency band. If, however, this fraction is as high as
suggested by the measurements (60--70\% for the total efficiency in
case of GRB~051221; \citealp{Soderbergetal06}) and GRB~051221 happened
indeed at a redshift of 0.546 and not farther away, then this
strong burst is also within the reach of the most energetic one of 
our models if observed near-axis.
Therefore we do not think that GRB~051221
gives strong support to speculations that different progenitors
and/or different energy extraction mechanisms than neutrino-antineutrino
annihilation are needed to explain the observed large spread in short
burst energies and the high energy of GRB~051221 in particular
\citep{Soderbergetal06}.

Because of the steep wings of the jet profiles in both
isotropic-equivalent energy and Lorentz factor $\Gamma$, our analysis
reveals a clear bias towards the detection of short GRBs with high
$\Gamma$ values. Since higher Lorentz factors and steep jet edges may
be characteristic features of GRB jets that originate from post-merger
BH-torus systems, which makes them different from collapsar jets, we propose
that they are the reason why short GRBs are typically harder than long
ones. In the internal shock scenario for GRB emission, this might imply
that the peak energy of the synchrotron spectrum is lower
($E_{\mathrm{p}} \propto R_{\mathrm{int}}^{-1}\propto 
\Gamma^{-2}\delta t^{-1}$). Short bursts
could therefore be harder not because of a higher $E_{\mathrm{p}}$
but because of a steeper increase of their $\nu F_\nu$ spectra below
the peak energy. This would require the spectral maximum to lie above or
inside the high-energy band for the hardness ratio.  Such properties
were indeed concluded from a spectral analysis of a sample of short
bright BATSE GRBs in comparison with long burst spectra (Ghirlanda
\etal\citeyear{GGC04}). Alternatively, a higher $\Gamma$ could still
allow for a larger $E_{\mathrm{p}}$ if the variability timescale
$\delta t$ of short GRBs were generically much smaller than that
of long GRBs. 

We point out that our exploration can only be tentative.  Both the
theoretical models and observational data still involve large
uncertainties, which have an impact on our analysis.  Not only did we
take a very simplistic approach in estimating the $\gamma$-ray
production of our jet models in the energy band of the detector.  We
just assumed that a fraction of 10\% of the ejecta energy is
converted to $\gamma$-emission at the relevant frequencies. We also
did not account in detail for the instrument specific detection
probability that depends on the photon flux and thus on the burst
luminosity, spectrum, and redshift, but simply assumed that all short
GRBs above a certain fluence threshold are detected, independent of
their peak luminosity and spectral properties. Moreover, the
intrinsic short GRB rate per unit of comoving time and comoving volume
enters the analysis sensitively. Nakar
\etal(\citeyear{NGF05}) found that the well-localized short GRBs,
which are all at $z < 1$, require a very large number of nearby 
events. This implies that the local number density of double neutron
star binaries or neutron star-black hole systems is significantly above
previous estimates, if short GRBs come from such objects. In fact,
it must be dominated by a so far undetected old population of double
neutron stars or, alternatively, by observationally unconstrained
NS+BH systems. A caveat of such conclusions and of comparisons to
the observations is, of course, the still very small sample of 
well-localized short GRBs. Caution is therefore advisable until the
empirical foundation is more solid (see \citealp{BP06}).

We restricted our analysis to the properties of the prompt emission
of short GRBs, which may originate from
neutrino-annihilation driven hydrodynamic jets, launched by the
neutrino energy release of a hyperaccreting BH girded by a massive 
torus. Of course, one cannot exclude that magnetohydrodynamic
effects might play a role that deserves more theoretical exploration.
The discovery of X-ray flares following the short-duration, hard 
pulse of GRB~050724 after hundreds of seconds (Barthelmy et al.\ 2005)
was interpreted as observational hint for the importance of such
magnetic processes (Fan, Zhang, \& Proga 2005). If these late-time
X-ray flares are indeed connected with an extended period of
activity or a restart of the central engine long after the prompt
emission is over (Burrows et al.\ 2005,
Zhang et al.\ 2005, Liang et al.\ 2006), it seems not possible
to explain these late outbursts by the accretion-rate dependent
neutrino mechanism (Fan, Zhang, \& Proga 2005). A prompt phase
of neutrino (or neutrino plus magnetic) energy release that 
creates the short-hard GRB may
thus be followed by a much more extended period of reduced source
activity, in which X-ray flares could be powered by    
a magnetic mechanism (Fan et al.\ 2005).

Admittedly, our investigation of the implications of the off-axis
properties of short GRB jets from post-merger BH-torus systems was
also based on a very limited set of numerical models (AJM05). This
theory input of our analysis replaces the intrinsic peak luminosity
function based on measured data used in other works 
(\eg Nakar \etal\citeyear{NGF05}, \citealp{GP05}). The models were 
chosen to span approximately the possible diversity of the properties of
ultrarelativistic outflows from such systems, but certainly represent
it only incompletely. In particular, they were calculated with a
single setup for the BH-torus system, in which the deposition rate of
thermal energy and its spatial distribution (expressed in terms of the
opening angle of the axial cone of the main energy deposition) was
varied over a range guided by the results for neutrino-antineutrino
annihilation in independent binary merger and post-merger evolution
models.  Neither was the torus evolved self-consistently in response
to this energy release and to the jet formation, nor were
magnetohydrodynamic effects included in the models.  There is still a
long way to go before the challenging problem of fully consistent
modeling becomes manageable. This should include all relevant physics
and track in full general relativity the history from the last
orbits of the pre-merging binary, through the merging phase, to the
possible BH formation and the subsequent secular post-merging
evolution of the accreting relic black hole. Such simulations will
ultimately have to be performed for a carefully selected sample of
progenitor systems, which avoids the ``arbitrariness'' of the
variations in our current set of models.  Knowing that this goal is
still far ahead and in view of the fact that the simulations available
to us represent the present state-of-the-art of modeling collimated
relativistic outflows from post-merger BH-torus systems, we tried to
explore how far this information can help us bridging
GRB engine theory and observations.

\begin{figure}
\plotone{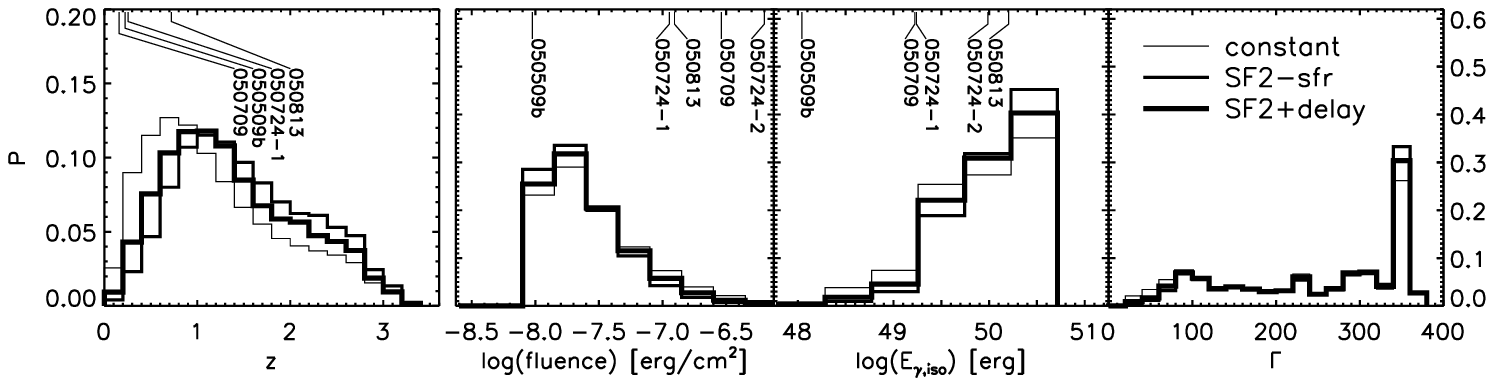}
\plotone{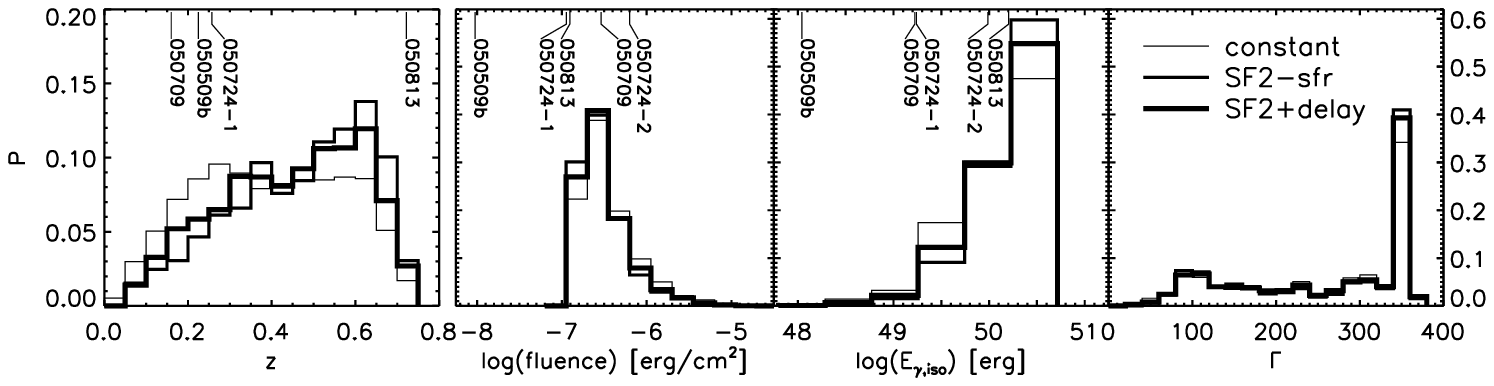}
 \caption{Expected normalized probability distributions of
   short GRBs as functions of redshift $z$,
   fluence, isotropic-equivalent energy output in $\gamma$-rays, 
   $E_{\gamma,\mathrm{iso}}$,
   and Lorentz factor $\Gamma$. The theoretical predictions are
   based on the results of our sample of jet models for 
   $E_{\gamma,\mathrm{iso}}$ and $\Gamma$ 
   shown in Fig.~\ref{fig:jetproperties2} and
   Fig.~\ref{fig:jetproperties}, respectively. In the {\em upper panels}
   we have adopted a lower fluence cutoff of $10^{-8}\,$erg$\,$cm$^{-2}$
   for the burst detectability, in the {\em bottom panels} we used 
   $1.6\times 10^{-7}\,$erg$\,$cm$^{-2}$ (note the corresponding
   change in the scales of the horizontal axes of the left two panels).
   The Lorentz factors $\Gamma$ shown in the right panels are the
   values from the hydrodynamic jet models at 0.5$\,$ after jet 
   formation. The terminal Lorentz factors $\Gamma_\infty$ will be
   significantly higher (see text).
   The lines correspond to three cases of different assumed
   comoving short GRB rate density as function of $z$ (see text
   for details). The locations of the observed short GRBs listed  
   in Table~1 are indicated. The very energetic GRB~051221 is outside
   of the energy scale of the third panel and therefore it is 
   omitted from all other panels of the figure as well.
\label{fig:grbobservability}}
\end{figure}

\acknowledgments{
  Stimulating discussions with Shri Kulkarni and Elena Rossi and help
  by Leonhard Scheck with Fig.~\ref{fig:grbobservability} are 
  acknowledged. We are very grateful to anonymous referees for their
  very interesting comments, which led to significant improvements of 
  our manuscript. The project was supported by SFB 375 ``Astroparticle
  Physics'' and SFB-TR 7 ``Gravitational Wave Astronomy''
  of the Deutsche Forschungsgemeinschaft, and by the RTN program 
  HPRN-CT-2002-00294 ``Gamma-Ray Bursts: an Enigma and a Tool''.
  MAA is a Ram\'on y Cajal
  Fellow of the Spanish Ministry of Education and Science and is
  grateful for partial support of the Spanish Ministerio de
  Ciencia y Tecnolog\'{\i}a (AYA2001-3490-C02-C01). 
  }

\clearpage
\begin{table*}
\begin{minipage}{160mm}
\scriptsize
\caption{Observed short GRBs with determined redshifts.}
  \tabcolsep=0.8mm
\begin{tabular}{@{}lccccccccccccccc@{}}
\\
\hline
GRB$^{\rm a}$ & $z$ & Band & $T_{\mathrm{ob}}^{\rm b}$ & 
$T_\mathrm{i}^{\rm c}$ & Fluence & $E_{\gamma,\mathrm{iso}}^{\rm d}$
& $\left\langle L\right\rangle^{\rm e}$ & $E_{\mathrm{p}}$ & $\eta^{\rm f}$
& $k^{\rm g}$ & Gal. & d$^{\rm h}$ & SFR & Ref.$^{\rm i}$ \\
&  & keV & ms & ms & $10^{-8}\,$erg$\,$cm$^{-2}$ & $10^{50}\,$erg & 
$10^{50}\,$erg$\,$s$^{-1}$ & keV & & & & kpc & 
$M_\odot\,{\mathrm{yr}}^{-1}$ & \\
\hline
  050509b & 0.225 & 15--150 & 40 & 33 & $0.95\pm 0.25$ 
  & 0.011 & 0.8--1.5 & $>$
  150 & $1.5 \pm 0.4$ & 2.3--4.3 & Ell & 40 & $< 0.1$ & 1,5 \\
  050709 & 0.16 &30--400 & 70 & 60 & $29 \pm 4$ & 0.167 & 4.7
  & $83^{+18}_{-12}$
  & $0.7 \pm 0.2$ & 1.7 & SF & 3.8 & 0.2 & 2,6 \\
  050724$^{\rm j}$ & 0.257 & 15--350 & 250 & 200 & $11.3 \pm
  8.7$ & 0.174 &
  1.4--1.7 & $>$350 & $1.7 \pm 0.2$ & 1.6--1.9 & Ell & 2.6 & $< 0.02$
  & 3,7 \\
  &&& 3000 & 2387 & $63 \pm 10$ & 0.97 & 0.7--0.8 & ... & & & &&& \\
  050813 & 0.722 & 15--350 & 600 & 350 & $12.4 \pm 4.6$ & 1.6 &
  7.8--13 & $>$350
  & $1.2 \pm 0.3$ & 1.7--2.8 & ? & -- & -- & 4,8 \\
  051221 & 0.546 & 20--2000 & 1400 & 906 & 320$^{+0.1}_{-1.7}$ & 24
  & 26 & $400\pm 80$ & -- & 1.0 & SF & 0.8 & 1.5 & 9 \\
  \hline 
   \multicolumn{15}{l}{$^{\rm a}$ All the events where
    detected by Swift except GRB~050709, which was observed by HETE-2.} \\
  \multicolumn{15}{l}{$^{\rm b}$ Observed duration $T_{90}$.} \\
  \multicolumn{15}{l}{$^{\rm c}$ Intrinsic duration 
           (cosmologically corrected).}\\
  \multicolumn{15}{l}{$^{\rm d}$ Isotropic equivalent gamma-ray energy 
           released by the burst in the specified band 
           (cosmologically corrected).}\\
  \multicolumn{15}{l}{$^{\rm e}$ Average luminosity (computed as
    $\left\langle L\right\rangle \equiv 
   k\,E_{\gamma,\mathrm{iso}}/T_\mathrm{i}$).} \\
  \multicolumn{15}{l}{$^{\rm f}$ Photon index of the average GRB
    spectrum ($f(E) \propto E^{-\eta}$).} \\
  \multicolumn{15}{l}{$^{\rm g}$ Bolometric corrections to
    rest-frame energies of 20-2000 keV, following \cite{Bloom01}.}\\
  \multicolumn{15}{l}{$^{\rm h}$ Projected distance of GRB from host center.}\\
  \multicolumn{15}{l}{$^{\rm i}$ References: 1. \cite{Gehrelsetal05}; 
   2. \cite{Boer05}; 3. \cite{Krimm05}; 4. \cite{Sato05};}\\
  \multicolumn{15}{l}{~~~ 5. \cite{Bloometal05}; 6. \cite{Foxetal05};
    7.  \cite{Bergeretal05}; 8. \cite{Foley05}; 
    9. \cite{Soderbergetal06}.}\\
  \multicolumn{15}{l}{$^{\rm j}$ This GRB shows a first short pulse,
    followed by a softer and longer event.  We report data for
    both events.}
\end{tabular}
\end{minipage}
\label{table:grbs}
\end{table*}

\clearpage
\end{document}